\documentclass[10pt,conference]{IEEEtran}
\IEEEoverridecommandlockouts
\usepackage{cite}
\usepackage{amsmath,amssymb,amsfonts}
\usepackage{url}
\usepackage{algorithmic}
\usepackage{graphicx}
\usepackage{textcomp}
\usepackage{multirow}
\usepackage{xcolor}
\usepackage{listings}
\usepackage{balance}
\usepackage{subcaption}
\usepackage[ruled,linesnumbered]{algorithm2e}
\def\BibTeX{{\rm B\kern-.05em{\sc i\kern-.025em b}\kern-.08em
    T\kern-.1667em\lower.7ex\hbox{E}\kern-.125emX}}

\usepackage{listings}
\definecolor{antlrmagenta}{rgb}{1, 0, 0.40}
\lstdefinestyle{ANTLR}{
    basicstyle=\small\ttfamily\color{antlrmagenta},%
    breaklines=true,%
    moredelim=[s][\color{green!50!black}\ttfamily]{`}{'},%
    moredelim=*[s][\color{black}\ttfamily]{options}{\}},%
    commentstyle={\color{gray}\itshape},%
    morecomment=[l]{//},%
    emph={%
        STRING, color, icon, meta, name, private, var, grammar, fragment%
        },emphstyle={\color{blue}\ttfamily},%
    alsoletter={:,|,;},%
    morekeywords={:,|,;},%
    keywordstyle={\color{black}},%
    	tabsize=3,%
    captionpos=b,%
    keepspaces,%
}
\definecolor{prismgreen}{rgb}{0, 0.5, 0} 
\lstdefinelanguage{Prism}{ %
        basicstyle=\color{red}\scriptsize\ttfamily, %
        breaklines=true,%
        keywords={bool,C,ceil,const,ctmc,double,dtmc,endinit,endmodule,endrewards, endsystem,F,false,floor,formula,G,global,I,init,int,label,max,mdp,min, module,nondeterministic,P,Pmin,Pmax,prob,probabilistic,R,rate,rewards, Rmin,Rmax,S,stochastic,system,true,U,X},%
        keywordstyle={\bfseries\color{black}},%
        numberstyle=\tiny\color{black},%
        comment=[l] {//}, morecomment=[s]{/*}{*/}, %
        commentstyle= \color{prismgreen}, %
        tabsize=2, %
        captionpos=b, %
        escapechar=@ %
}

\usepackage{pgfplots}
\usepackage{pgfplotstable}
\pgfplotsset{compat=newest}
\pgfplotscreateplotcyclelist{my chart colors}{%
blue, solid, every mark/.append style={solid, fill=blue}, mark=*\\%
red, solid, every mark/.append style={solid, fill=red}, mark=square*\\%
brown, solid, every mark/.append style={solid, fill=brown}, mark=triangle*\\%
cyan, solid, every mark/.append style={solid, fill=gray}, mark=diamond*\\%
black, dashdotted, every mark/.append style={solid, fill=red}, mark=halfdiamond*\\%
olive, dashdotted, every mark/.append style={solid, fill=green}, mark=halfcircle*\\%
orange, dashdotted, every mark/.append style={solid, fill=gray}, mark=*\\%
purple, dashdotted, every mark/.append style={solid, fill=blue}, mark=halfsquare left*\\%
teal, densely dashed, every mark/.append style={solid, fill=green}, mark=oplus*\\%
violet, densely dashed, every mark/.append style={solid, fill=gray}, mark=diamond*\\%
magenta, densely dashed, every mark/.append style={solid, fill=blue}, mark=triangle*\\%
lime, densely dashed, every mark/.append style={solid, fill=red}, mark=square*\\%
}

\usepackage{graphicx}

\renewcommand{\baselinestretch}{0.94}
\graphicspath{{figures/}}

\usepackage[normalem]{ulem}
\usepackage{xcolor}

\makeatletter
\def\ps@IEEEtitlepagestyle{
	\def\@oddfoot{\mycopyrightnotice}
	\def\@evenfoot{}
}
\def\mycopyrightnotice{
	{\footnotesize
		\begin{minipage}{\textwidth}
			\centering
			\textcopyright~2019 IEEE. Personal use of this material is permitted.  Permission from IEEE must be obtained for all other uses, in any current or future media, including reprinting/republishing this material for advertising or promotional purposes, creating new collective works, for resale or redistribution to servers or lists, or reuse of any copyrighted component of this work in other works.
		\end{minipage}
	}
}
\usepackage[hidelinks]{hyperref}
\hypersetup{
	bookmarks=true,         
	unicode=false,          
	pdftoolbar=true,        
	pdfmenubar=true,        
	pdffitwindow=false,     
	pdftitle={Taming Uncertainty in the Assurance Process of Self-Adaptive Systems: a Goal-Oriented Approach},
	pdfauthor={Gabriela Felix Solano, Ricardo Diniz Caldas, Genaina Nunes Rodrigues, Thomas Vogel, Patrizio Pelliccione},
	pdfsubject={},
	pdfcreator={},
	pdfproducer={},
	pdfkeywords={Self-adaptive systems, uncertainty, goal modeling, symbolic model checking, adaptation policy},
	pdfnewwindow=true,
}

\begin{document}

\title{Taming Uncertainty in the Assurance Process of Self-Adaptive Systems: a Goal-Oriented Approach}

\author{\IEEEauthorblockN{Gabriela F\'elix Solano,\\Ricardo Diniz Caldas,\\Gena\'ina Nunes Rodrigues}
\IEEEauthorblockA{University of Bras\'ilia\\
Bras\'ilia, DF, Brazil\\
felix.solano@gmail.com,\\rdinizcal@gmail.com,genaina@unb.br}
\and
\IEEEauthorblockN{Thomas Vogel}
\IEEEauthorblockA{Humboldt-Universit\"at zu Berlin\\
Berlin, Germany\\
thomas.vogel@cs.hu-berlin.de}
\and
\IEEEauthorblockN{Patrizio Pelliccione}
\IEEEauthorblockA{Chalmers \textbar~University of Gothenburg\\
Gothenburg, Sweden\\
University of L'Aquila\\
L'Aquila, Italy\\
patrizio@chalmers.se}}

\maketitle

\begin{abstract}
Goals are first-class entities in a self-adaptive system (SAS) as they guide the self-adaptation. A SAS often operates in dynamic and partially unknown environments, which cause uncertainty that the SAS has to address to achieve its goals. Moreover, besides the environment, other classes of uncertainty have been identified. However, these various classes and their sources are not systematically addressed by current approaches throughout the life cycle of the SAS. In general, uncertainty typically makes the assurance provision of SAS goals exclusively at design time not viable. This calls for an assurance process that spans the whole life cycle of the SAS.
In this work, we propose a goal-oriented assurance process that supports taming different sources (within different classes) of uncertainty from defining the goals at design time to performing self-adaptation at runtime. 
Based on a goal model augmented with uncertainty annotations, we automatically generate parametric symbolic formulae with parameterized uncertainties at design time using symbolic model checking. These formulae and the goal model guide the synthesis of adaptation policies by engineers. At runtime, the generated formulae are evaluated to resolve the uncertainty and to steer the self-adaptation using the policies. 
In this paper, we focus on reliability and cost properties, for which we evaluate our approach on the Body Sensor Network (BSN) implemented in OpenDaVINCI.
The results of the validation are promising and show that our approach is able to systematically tame multiple classes of uncertainty, and that it is effective and efficient in providing assurances for the goals of self-adaptive systems.
\end{abstract}


\section{Introduction}
\label{sec:introduction}

Uncertainty in a software system is defined as the circumstances in which the system's behavior deviates from expectations due to dynamicity and unpredictability of a variety of factors existing in such systems~\cite{mahdavi2017classification}. For this reason, uncertainty is a fundamental challenge for a self-adaptive system (SAS) and pervades from the system's requirements to its infrastructure and runtime environment~\cite{esfahani2013uncertainty}. Many sources of uncertainty have been identified and further grouped to four classes: (i)~the system itself, (ii)~the system goals, (iii)~the environment, and (iv)~human aspects~\cite{mahdavi2017classification, weyns2017software}.  
These various classes of uncertainty have to be addressed by SAS assurance processes to comprehensively and systematically tame %
(address) such uncertainties. Otherwise, the provided assurances may render themselves inaccurate or incomplete.
Therefore, uncertainty should be leveraged as a first-class concept in self-adaptation~\cite{esfahani2013uncertainty} and in the assurance process~\cite{weyns2017software}. 
We generally consider an assurance process as all design-time and runtime activities providing evidence that the adaptation goals are satisfied. 

However, a SAS faces a paradoxical challenge in the presence of uncertainties~\cite{weyns2017software}: ``{\em how can one provide guarantees for the goals of a system that is exposed to continuous uncertainties?}''. There has been extensive research to address uncertainty in SAS~\cite{mahdavi2017classification, de2017software}, but with no focus on proposing solutions to systematically tackle classes of uncertainty and its sources\cite{mahdavi2017classification}. Also, the main focus has been on environmental uncertainty (regarding the variability of execution contexts), and on system goals uncertainty (regarding goal changes)~\cite{mahdavi2017classification}. Moreover, most of the existing approaches postpone the treatment of uncertainty to the runtime phase of the system's life cycle~\cite{mahdavi2017classification} while rather neglecting the explicit management of uncertainty right from the start with the requirements.
For instance, we proposed in previous work a goal-based feedback loop for real-time SAS that enhances the assurance process of goals verified offline with online learning~\cite{Rodrigues:learning}. The focus has been on real-time properties and learning, without explicitly managing (any class of) uncertainty as a first-class concept. 

To recognize and manage uncertainties in the assurance process from early on, we propose an end-to-end goal-oriented approach based on Goal-Oriented Requirements Engineering (GORE)~\cite{Lamsweerde:2001}. GORE offers proved means to decompose technical and non-technical requirements into well-defined entities (goals) and reason about the alternatives to meet them. Hence, it has been used as a means to model and reason about the systems' ability to adapt to changes in dynamic environments~\cite{ali2010goal, Lamsweerde:2001, Mendonca:2014, whittle2009relax}. Based on GORE, our approach leverages goal modeling to support the modeling of different sources of uncertainty within the following classes: (1)~system itself, (2)~system goals, and (3)~environment. We should note that by system we mean the managed system that comprises application code to realize domain functionality~\cite{weyns2017software}. The approach further supports the automatic generation of trustworthy verifiable models \emph{parameterized with uncertainties}, which are used at runtime to assist the assurance of a SAS.

Our assurance process to tame uncertainty aims at both design- and runtime of SAS. At design time, we augment goal modeling with uncertainties in which the leaf tasks represent executable components in the system and have their respective reliability, cost, and frequency (w.r.t. usage profiles) properties. Then, we automatically translate the resulting goal models into reliability and cost parametric formulae using symbolic model checking~\cite{daws2004symbolic}. These formulae are used as runtime models to express probabilities over the fulfillment of SAS goals. Specifically, they take different classes of uncertainty into account while supporting self-adaptation. To overcome the state-space explosion problem of traditional model checking~\cite{kwiatkowska2006quantitative}, we compositionally generate such parametric symbolic formulae from Markov Decision Processes (MDPs) with parameterized uncertainties based on our augmented goal model. At runtime, based on the idea of feedback control~\cite{astrom2010feedback, hellerstein2004feedback}, the controller (henceforth called \emph{managing system}~\cite{weyns2017software}) continuously monitors the costs and reliability statuses of the managed system as well as context conditions to resolve the parameterized uncertainties. Then, the parametric (runtime) models are exercised to (i)~evaluate the system's reliability and cost, and (ii)~evaluate policy actions that should be triggered to achieve the goals, thus guiding adaptation decisions in SAS.

We evaluate our work by the ability of our assurance process to efficiently support the managing system in taming uncertainties.
For this purpose, we extended our previous Body Sensor Network~(BSN) artifact~\cite{Rodrigues:learning} with new features to incorporate uncertainties that could hinder reliability and cost properties of patients assisted by the BSN. The adaptation policies synthesized by our approach for the BSN then tame such uncertainties by means of parametric symbolic formulae for reliability and cost.  Through our approach, the managing system is indeed able to adjust the managed system to work in the set point boundaries while taming different classes of uncertainties with significant reliability and cost trade-off enhancements (at least twice as much) over self-adaptation without taming uncertainties.
Moreover, our results show that runtime parametric model checking for the BSN is affordable (i.e., formula verification time is 0.02s, formula size is 22kB).

In a nutshell, to tame multiple classes of uncertainties, we propose an augmented assurance process of SAS that blends over the high-level representation of goals and low-level representation of synthesis and verification models of SAS through symbolic model checking with parameterized uncertainties. The contributions of this work are threefold: 

\begin{enumerate}
\item %
We introduce annotations to semantically enhance goal models with different classes of uncertainty, thus \emph{providing first class support for uncertainty} in goal modeling.
\item %
We develop an algorithm to automatically generate the parametric symbolic formulae from our enhanced goal model. These formulae containing parameterized uncertainties guide (i)~the synthesis of adaptation policies by engineers, and (ii)~the self-adaptation at runtime.
\item %
We provide two artifacts by (i)~extending the goal-oriented dependability analysis framework (GODA)~\cite{mendoncca2016goda} with the generation of parametric symbolic formulae with parameterized uncertainties, and (ii)~evaluating our approach on an extended version of the BSN~\cite{Rodrigues:learning} to validate the assurance process we propose in this work.
\end{enumerate}

The rest of the paper is structured as follows. In Section~\ref{sec:background}, we provide background knowledge on the extended Body Sensor Network (BSN) used as our running example. We discuss the proposed approach in Section~\ref{sec:approach} and evaluate our proposal in Section~\ref{sec:experiments}. Section~\ref{sec:relatedwork} highlights related work and Section~\ref{sec:conclusion} concludes the paper along with future work.

\section{The Body Sensor Network Example}
\label{sec:background}

To illustrate our approach, we use the example of a Body Sensor Network~(BSN)~\cite{pessoa2017building} enriched with new features to incorporate uncertainties that affect reliability and cost (e.g., power consumption) of the BSN. The main objective of the BSN is to keep track of a patient's health status, continuously classifying it into \emph{low}, \emph{normal}, or \emph{high risk}, and to send an emergency signal to a central unit in the case of an anomaly. The structure of the BSN is as follows: several wireless sensors are connected to a person to monitor her vital signs, namely, an electrocardiograph sensor (ECG) for heart rate beats measurement, a pulse oximeter for blood oxygen saturation (SaO2), a thermometer for body temperature (TEMP) in Celsius, and a sphygmomanometer for measuring diastolic and systolic arterial blood pressure (ABP). Additionally, the central node can preprocess the collected data, filter redundancy, translate communication protocols, and fuse data. Table~\ref{table:operat} shows how the sensor values and thus, how the context relates to the patient's health risk as specified by a domain expert. 
The implementation of BSN is based on OpenDaVINCI~\cite{opendavinci}.

\begin{table}[b]
\renewcommand{\arraystretch}{1.2}
    \caption{Context Operationalization for a Patient's Status.}
    \label{table:operat}
    \centering
    \scalebox{0.97}{
    \begin{tabular}{ll}
    \hline
        \multicolumn{1}{c}{\textbf{Sensor Info}} & \multicolumn{1}{c}{\textbf{Data Ranges}}
        \\\hline
        Oxygen saturation:      & 100 $>$ low $>$ 65 $>$ medium $>$ 55 $>$ high $>$ 0                           \\\hline
        Heart beats:       & \begin{tabular}[c]{@{}l@{}} 300 $>$ high $>$ 115 $>$ medium $>$ 97 $>$ low $>$ \\ 85 $>$ medium $>$ 70 $>$ high $>$ 0   \end{tabular}                        \\\hline
        Temperature:     & \begin{tabular}[c]{@{}l@{}} 50 $>$ high $>$ 41 $>$ medium $>$ 38 $>$ low $>$    \\
         36 $>$ medium $>$ 32 $>$ high $>$ 0
        \end{tabular} 
         \\\hline
         Systolic Pressure:     & \begin{tabular}[c]{@{}l@{}} 300 $>$ high $>$ 140 $>$ medium $>$ 120 $>$ low $>$    0
        \end{tabular} 
         \\\hline
         Dyastolic Pressure:     & \begin{tabular}[c]{@{}l@{}} 300 $>$ high $>$ 90 $>$ medium $>$ 80 $>$ low $>$     0 
        \end{tabular} 
         \\\hline
    \end{tabular}
    }
\end{table}

In general, a Contextual Goal Model (CGM) is composed of actors, goals, tasks, and contexts~\cite{ali2010goal}. 
\textit{Actors} are humans or software with the ability to decide autonomously on how to achieve their goals. \textit{Goals} are abstractions to represent stakeholders' needs and expectations, which offer an intuitive way to elicit and analyze requirements, while \textit{tasks} are responsible for the operationalization of a these goals, that is, an operational means to reach them. A goal is satisfied by the compositional fulfillment of its tasks. Finally, \textit{contexts} are partial states of the world that are relevant to a goal as context changes may influence the (quality of the) goals and the means of achieving them. Goals and tasks of a CGM can be refined into AND-decomposition (OR-decomposition), that is, a link that decomposes a goal/task into subgoals/subtasks, meaning that all (at least one) of the subgoals/subtasks must be fulfilled/executed to satisfy its parent entity. The link between a goal and a task is called means-end, and indicates a means to fulfill a goal through the execution of a task.

\begin{figure}[t]
\centerline{\includegraphics[width=\columnwidth]{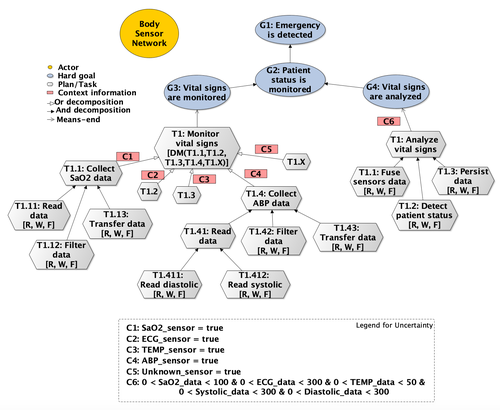}}
\caption{Contextual Goal Model of the Body Sensor Network.}
\label{fig:ND-CGM}
\vspace{-1em}
\end{figure}

Fig.~\ref{fig:ND-CGM} shows an excerpt of the CGM for the BSN. 
The root goal of the actor ``Body Sensor Network'' is ``G1:~Emergency is detected", which is refined into ``G2:~Patient status is monitored". G2 is divided into two subgoals: ``G3:~Vital signs are monitored" and ``G4:~Vital signs are analyzed''. Such goals are realized by tasks that are refined to leaf tasks that are operational. For instance, G3 is realized by ``T1:~Monitor vital signs'' that is refined to ``T1.1:~Collect SaO2 data'', 
``T1.2:~Collect ECG data'', 
``T1.3:~Collect temperature data'', 
``T1.4:~Collect ABP data'',
and
``T1.X''. 
The tasks T1.2 and T1.3 are not detailed in Fig.~\ref{fig:ND-CGM} but their further refinements to leaf tasks are analogous to task T1.1 and T1.4. 
The placeholder task T1.X exist because of uncertainty and will be discussed in the following section.
The operation of the BSN is subject to six different context conditions. Contexts C1 to C5 refer to the availability of sensors that collect data related to a patient's vital signs, which may be \textit{true} or \textit{false}. Context C6 refers to the operationalization of collected data, stipulating ranges of valid integer or double values that each sensed data can assume when analyzing a patient's health status for risks~\cite{pessoa2017building}. %

Contexts C1 to C5 cause a non-deterministic behavior such that the fulfillment of G3 depends on the execution of any combination of tasks T1.1, T1.2, T1.3, T1.4, and T1.X. The behavior assumed at runtime is related to the different conjunction of contexts that currently hold, and each conjunction shapes the system to fulfill a goal at a different quality level.

\section{A Goal-Oriented Approach to Tame Uncertainties in SAS}
\label{sec:approach}

\begin{figure}[h!]
\vspace{-1.3em}
\centerline{\includegraphics[width=\columnwidth]{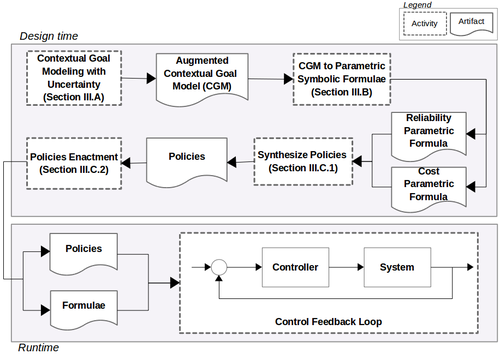}}
\caption{Assurance Process to Tame Uncertainty.}
\label{fig:approach}
\vspace{-.5em}
\end{figure}

Our approach aims at trustworthy SAS. It relies on the model representation of a SAS under uncertainty that allows a reasoned and precise analysis of reliability and cost properties and that supports the synthesis of adaptation policies.
Fig.~\ref{fig:approach} provides an overview of our approach with its four activities:

\textit{(1)~Contextual Goal Modeling with Uncertainty}: 
At design time, our approach extends the GODA framework~\cite{mendoncca2016goda} to support goal modeling that takes different classes of uncertainty into account. This results in augmented contextual goal models (CGMs) that make the uncertainty explicit.

\textit{(2)~CGM to Parametric Symbolic Formulae}:
Using the augmented CGM, our approach automatically generates formal models for future analysis, particularly the formulae with parameterized uncertainties targeting reliability and cost properties. These formulae keep the uncertainty explicit.

\textit{(3)~Synthesize Policies}:
The extended CGM and the formulae are then used to guide engineers in synthesizing adaptation policies. For this purpose, the model and formulae inform the engineers about the goals, contexts, and uncertainty that have to be taken into account.

\textit{(4)~Policy Enactment}:
Finally, a subset of eligible policies are enacted at runtime, which will guide the adaptation of the managed system to assure the satisfaction of goals supported by the reliability and cost parametric formulae. %

Although the contributions of the approach mainly refer to the design-time phase, we have integrated and evaluated the policies and formulae in a feedback loop for self-adaptation. We provide an interface between the formulae and the feedback loop so that the analysis of the managed system as well as the evaluation and selection of policies for self-adaptation are based on these formulae instead of complex runtime models as proposed and used in~\cite{Vogel:2010, Bennaceur:2014,blair2009models}. 

In the remainder of this section, we detail the approach and discuss each of its activities shown in Fig.~\ref{fig:approach}.

\subsection{Contextual Goal Modeling with Uncertainty}
\label{subsec:goalModeling}

The first step of our approach consists of designing a goal model of a SAS that takes uncertainty into account\footnote{The implementation is available at GitHub \url{https://github.com/gabrielasolano/Seams2019}. The modeling and analyzing environment of our GODA extension is available at Heroku \url{https://seams2019.herokuapp.com/}.}. For this purpose, we augment CGM and introduce annotations to support three classes of uncertainty: (1)~the system itself, (2)~system goals, and (3)~the environment, as well as non-deterministic behavior. 
According to Mahdavi-Hezavehi~\textit{et~al.}~\cite{mahdavi2017classification}, uncertainty related to the system itself refers to the managed system (i.e., the subsystem under adaptation); uncertainty related to system goals refers to the specification, modeling, and alteration of goals; and uncertainty related to the environment refers to environmental circumstances that interact with or affect the system.

In our work, each executable task (i.e., leaf task) of the CGM is implemented by a component in the managed system that has an identification label, description, execution probability (or equivalently, service usage profile), execution cost, and reliability. In particular, the execution probability indicates the execution profile over time, the cost indicates the energy consumption, and the reliability represents the probability of a successful execution of the component. 
Thus, the mapping of leaf tasks to components connects the CGM to the managed system.
On the other hand, the context variable in the CGM is an outcome of the context operationalization process~\cite{arthur2018,knauss2016acon} that has an identification label, description, and Boolean value that indicates whether a context is currently active or not, thus connecting the CGM to the managed system's environment.

In the following, we discuss the CGM extensions for uncertainty of the system itself, system goals, and environment, as well as for modeling non-deterministic behavior.

\noindent\textbf{System itself.}
Regarding uncertainty related to the system itself, we focus on two sources: \emph{future parameter value}, and \emph{incompleteness}~\cite{mahdavi2017classification}. To represent a future parameter value in the CGM, that is, uncertainty of system values such as reliability and cost of components executing tasks, we associate corresponding parameters to each executable task. Thus, the reliability and cost of a component are annotated to the leaf task the component implements. In Fig.~\ref{fig:ND-CGM}, these parameters are the annotation \textit{[R,W]} to each leaf task, where \textit{R} refers to reliability and \textit{W} to cost. We should note that such parameters are intrinsic to each leaf task in our augmented CGM so that they are not required to be explicitly attributed to any particular value and neither explicitly modeled. In contrast, the values will be obtained by monitoring the execution of the managed system's components.
The \emph{incompleteness} of the system is represented by a placeholder node ``X'' for a goal or task 
that is not entirely known at design-time. A replacement for such a placeholder is only known at runtime. For instance, task ``T1.X'' in Fig.~\ref{fig:ND-CGM} will be replaced at runtime by some data collection task that is unknown at design-time. 
Despite we model the incompleteness of the system as an uncertainty in the CGM, we leave its taming for future work.

\noindent\textbf{System goals.}
The support for uncertainty related to system goals focuses on the \emph{specification of goals}~\cite{mahdavi2017classification}, specifically on enhancing the accuracy of the specification of stakeholders' preferences~\cite{weyns2017software}. To this end, we associate leaf tasks with their frequencies of execution. This frequency represents the usage profile of the corresponding task so that it describes an external view of the use of each component of the managed system. 
Similar to the reliability and cost properties, the frequency is associated to leaf tasks as shown in Fig.~\ref{fig:ND-CGM}. The annotation \textit{[F]} to each leaf task represents the task's execution frequency. Similar to the annotation \textit{[R,W]} %
for the reliability and cost parameters of a task, the parameter \textit{F} is intrinsic to each task. Therefore, it is not explicitly attributed to any particular value in the model and neither explicitly modeled. In contrast, its value will be obtained by monitoring the frequency of executing the managed system's components.

\noindent\textbf{Environment.}
Regarding uncertainty related to the environment, our approach supports \emph{execution context} and \emph{noise in sensing}~\cite{mahdavi2017classification}. The \emph{execution context} is inserted in the CGM as context annotations. Such annotations are associated with a node $n$ and they can assume \textit{integer}, \textit{double}, or \textit{Boolean} value types. In Fig.~\ref{fig:ND-CGM}, C1 to C6 represent the contexts in which the BSN operates and that needs to be monitored to determine a suitable behavior of the BSN during execution. 
To support \emph{noise in sensing}, we acknowledge the use of sensors by SAS (e.g., cyber-physical SAS) in its monitoring infrastructure to measure events in its environment~\cite{ramirez2012taxonomy}. We define a modeling guideline to take into account the data collection behavior of sensors. 
In this sense, a sensor's task of collecting data is decomposed into three essential sub-tasks.
The first task, \textit{read data}, is responsible for gathering data from the environment.
The second task, \textit{filter data}, is responsible for filtering and removing noise in the gathered data.
Finally, the third task, \textit{transfer data}, sends the filtered data to the managing system. 
For instance, the tasks ``T1.1:~Collect SaO2 data" and ``T1.4:~Collect ABP data" exemplify the use of our guideline for the blood oxygen saturation and arterial blood pressure sensors, respectively (see Fig.~\ref{fig:ND-CGM}). 
With this guideline, we support the use of sensors in SAS while mitigating the \emph{noise in sensing} associated with them through dedicated tasks.

\noindent\textbf{Non-deterministic behavior.}
Another extension of the CGM is the possibility of defining a system's non-deterministic behavior using the \textit{decision-making (DM)} annotation. Non-determinism in SAS is brought by uncertainty, such that the lack of runtime knowledge implicates planning multiple alternative behaviors at design time to accommodate possible configurations the system may face during execution.
The DM-annotation is only inserted in non-leaf nodes (goals or tasks), and it requires the node to be refined into context-dependent subtrees by an OR-decomposition. 
For instance in Fig.~\ref{fig:ND-CGM}, task ``T1: Monitor vital signs" has the annotation $DM(T_{1.1},T_{1.2},T_{1.3},T_{1.4},T_{1.X})$, which specifies that the successful execution of any combination of the context-dependent subtrees $T_{1.1}$, $T_{1.2}$, $T_{1.3}$, $T_{1.4}$, or $T_{1.X}$ will result in the successful fulfillment of T1. In other words, T1 has up to $2^5$ different ways to be satisfied, but the decision of which way to pursue is only made at runtime, after mapping each context of T1's subtrees to a value and eliciting the viable paths. 
For example, assuming that BSN is only operating with SaO2 and ECG sensors at a given time. This means that only contexts C1 and C2 are evaluated to \textit{true}, therefore only tasks $T_{1.1}$ and $T_{1.2}$ may fulfill their parent node T1 (see Fig.~\ref{fig:ND-CGM}). In this sense, the viable paths to fulfill T1 are: (i) through only $T_{1.1}$; (ii) through only $T_{1.2}$; or (iii) through both $T_{1.1}$ and $T_{1.2}$. By providing this annotation, our work avoids the enumeration of all possible paths brought by context variability, 
therefore avoiding a state space explosion problem in the goal model.

\subsection{CGM to Parametric Symbolic Formulae}\label{subsec:FromGMtoPF}

After the goal modeling, our approach automatically translates the augmented CGM to verifiable models, which avoids any overhead and errors caused by a manual generation step.

The generation of parametric formulae is a key enabler towards an affordable time-space runtime analysis~\cite{mendoncca2016goda}. Because of uncertainty and the corresponding non-deterministic behavior, model checking requires a Markov Decision Process (MDP)~\cite{kwiatkowska2013automated} to provide best- or worst-case scenarios analysis based on lower or upper probability bounds that can be guaranteed, when ranging over all possible paths~\cite{baier2008principles}. 
Consequently, symbolic model checking of a whole CGM with uncertainty through a single MDP may present memory and time limitations due to the quantity of parameters involved. Our approach overcomes these limitations by accounting for uncertainty and by allowing the model checking of independent and smaller parts of the system, thus compositionally generating unique formulae and avoiding combinatorial state explosion problems.

We build the parametric MDP (in PRISM language~\cite{prismlanguage}) for each leaf task (i.e., executable task) in the augmented CGM, and compose them to represent the fulfillment of the modeled goals. Listing~\ref{lst:PRISM_EXAMPLE} presents the PRISM template for a context-dependent leaf task $N1$. In particular, $c1$ is a binary placeholder that assumes value 1 (true) when the \emph{execution context} in the leaf task $N1$ holds, and value 0 (false) otherwise; $r1$ represents the \emph{reliability}, and $f1$ the \emph{execution frequency} of the task.
The latest two represent the success probability of a task and the usage percentage of a task in a given time, respectively. Therefore, they assume real values in the range of 0 to 1. 
Moreover, $s1$ models the state of the leaf task. State \textbf{init} (s1 = 0) corresponds to the initial state. Then, a leaf task can enter state \textbf{running} (s1 = 1) if it is selected to start execution, or the final state \textbf{skipped} (s1 = 3) if it does not participate in the fulfillment of the parent goal. Before, it is verified whether the context condition of a task, if existent, is satisfied or not. The probability of moving from \textbf{init} (s1 = 0) to \textbf{running} state (s1 = 1) is $c1 * f1$ if the leaf task is context dependent, and $f1$ otherwise. Consequently, $1 - (c1*f1)$ represents the probability of moving from \textbf{init} (s1 = 0) to \textbf{skipped} state (s1 = 3) for a context-dependent leaf task, and $1 - f1$ for a context-free leaf task. Once started running, the result of the task fulfillment is \textbf{success} (s = 2) with probability $r$, if the task has successfully been executed, otherwise \textbf{failure} (s = 4) with probability $1-r$. The \emph{cost} of executing the task is represented by $w1$ in the reward structure. Label $next$ is used to tag a transition that needs to be synchronized with other transitions in the same or in different modules, while variable $x$ represents an index that sequentially sets the actions of the model. The MDP for a goal $G_i$ is obtained by composing the MDP models of its subtrees.

\begin{lstlisting}[language=Prism, caption={CGM Context-dependent Leaf Task in PRISM Language.},label={lst:PRISM_EXAMPLE}]
const int c1; //context condition of n1
const double r1; //n1 probability of success
const double f1; //n1 frequency of execution
module N1
  s1 :[0..4] init 0;
  //init to running or to skipped
  [next<x>] s1 = 0 -> c1*f1 : (s1'=1)+(1-c1*f1) : (s1'=3);
  [] s1 = 1 -> r1 : (s1'=2) + (1 - r1) : (s1'=4); //running to final state
  [next<x+1>] s1 = 2 -> (s1'=2); //final state success
  [next<x+1>] s1 = 3 -> (s1'=3); //final state skipped
  [next<x+1>] s1 = 4 -> (s1'=4); //final state failure
endmodule
rewards "cost"
  s1 = 1 : w1; //cost of n1 execution
endrewards
\end{lstlisting}

\begin{lstlisting}[language=Prism, caption={CGM Node with a DM-annotation in PRISM Language.},label={lst:PRISM_MDP}]
const int CTX_1; //context condition of n1
const int CTX_2; //context condition of n2
const int CTX_3; //context condition of n1 & n2

global c1: [0..1] init 0; //variable that enables n1
global c2: [0..1] init 0; //variable that enables n2
module NonDeterminism
	s :[0..5] init 0;
	[next<x>] s = 0 -> (s'=1);
	[] s = 1 -> CTX_1 : (s'=2) + (1 - CTX_1) : (s'=1);
	[] s = 1 -> CTX_2 : (s'=3) + (1 - CTX_2) : (s'=1);
	[] s = 1 -> CTX_3 : (s'=4) + (1 - CTX_3) : (s'=1);
	[] s = 1 -> (s'=5); //no uncertainty holding
	//enable the correspondent tasks
	[] s = 2 -> (s'=5) & (c1'=1);
	[] s = 3 -> (s'=5) & (c2'=1);
	[] s = 4 -> (s'=5) & (c1'=1) & (c2'=1);
	[next<x+1>] s = 5 -> (s'=5);
endmodule
\end{lstlisting}

To compose the MDP model for a CGM node with a DM-annotation, we also model the non-determinism in PRISM as shown in Listing~\ref{lst:PRISM_MDP}. The $CTX$ parameters represent each combination of context conditions modeled in the node's subtrees. They assume integer values of 1 or 0 to indicate whether the represented context conditions are satisfied or not. Lines 10--13 present the non-determinism in the model that depends on the values of the $CTX$ parameters to be resolved. Once the non-determinism is resolved, the global variables $c1$ and $c2$ are either set to 1, enabling the execution of its corresponding leaf task, or to 0, indicating that the corresponding leaf task will enter skipped state. 

In this work, we focus on reliability and cost properties. Their specifications are based on the idea of the \textit{probabilistic existence property}~\cite{Grunske:2011}, that is, the probability that a system will \emph{eventually} reach a state that satisfies a goal of interest. The maximum and minimum reliability and cost of fulfilling a goal $G_i$ is defined in Table~\ref{tab:pctlPrismVerification}, where proposition~$\phi$ represents the success of $G_i$ and $\phi$ is recursively formed by composing the propositions of the nodes underlying $G_i$ in the CGM.

\begin{table}[htb]
\renewcommand{\arraystretch}{1.4}
\caption{Reliability and Cost Properties for Verification.}
\centering
\begin{tabular}{@{}ll@{}}
\hline
\textbf{Property} & \textbf{PCTL formula}\\
\hline
Reliability (max) & $Pmax_{G_i}=?\ [F\ (\phi)]$\\
Reliability (min) & $Pmin_{G_i}=?\ [F\ (\phi)]$\\
Cost (max) & $Cmax_{G_i}=?\ [F\ (\phi)]$\\
Cost (min) & $Cmin_{G_i}=?\ [F\ (\phi)]$\\
\hline
\end{tabular}
\label{tab:pctlPrismVerification}
\end{table}

Proposition $\phi$ varies for each node in the CGM according to the node's features: AND/OR-decomposition, DM-annotation, and incompleteness. Table~\ref{tab:phi} specifies the proposition for each feature, 
where $i \neq j$ are nodes in the CGM, and $C_i$ is the context information constraining node $i$. Note that a node specifying \emph{incompleteness} in the system is optional to the system overall fulfillment, since it depends on the availability of an unknown runtime resource. Note also that the proposition for a node with DM-annotation takes into account the \emph{execution context} of its subtree, since it is a mandatory information when using the DM-annotation. Other node $i$, with a different feature, that is also constrained by a context condition has its proposition defined as $\phi' = (!C_i \wedge skipped_i \vee \phi)$, where $\phi$ is also described according to Table~\ref{tab:phi}.

\begin{table}[htb]
\renewcommand{\arraystretch}{1.4}
\caption{Proposition of Success of a Node in the CGM.}
\centering
\begin{tabular}{@{}ll@{}}
\hline
\textbf{Node features} & \textbf{Proposition} \\
\hline
AND-decomposition & $\phi = succeeded_i \wedge succeeded_j$\\
OR-decomposition & $\phi = succeeded_i \vee succeeded_j$\\
DM-annotation & $\phi = succeeded_i \vee (!C_i \wedge skipped_i)$\\
Incompleteness & $\phi = succeeded_i \vee skipped_i$\\
\hline
\end{tabular}
\label{tab:phi}
\end{table}

\begin{table*}[htb]
\renewcommand{\arraystretch}{1.4}
\caption{Symbolic Formulae.}
\centering
\begin{tabular}{@{}lll@{}}
\hline
\textbf{Node features} & \textbf{Reliability formula} & \textbf{Cost formula} \\
\hline
AND $(N_1,N_2)$ & $C_{n1}P_{n1}*C_{n2}P_{n2}$ & $(C_{n1}W_{n1} + C_{n2}W_{n2})*P_{AND(n1,n2)}$\\
OR $(N_1,N_2)$ & $-C_{n1}P_{n1}*C_{n2}P_{n2}+C_{n1}P_{n1}+C_{n2}P_{n2}$ & $(C_{n1}W_{n1} + C_{n2}W_{n2})*P_{OR(n1,n2)} - C_{n2}W_{n2}*C_{n1}P_{n1}$ \\
$DM(N_1,N_2)$ & $-C_{n1}P_{n1}*C_{n2}P_{n2}+C_{n1}P_{n1}+C_{n2}P_{n2}$& $(C_{n1}W_{n1} + C_{n2}W_{n2})*P_{DM(n1,n2)} - C_{n2}W_{n2}*C_{n1}P_{n1}$ \\
Incompleteness $(N_x)$ & $C_x * P_x * OPT_x$& $C_x*W_x*P_x*OPT_x$ \\
\hline
\end{tabular}
\label{paramTable}
\end{table*}

With the MDP and property formulae for each node feature, we use the symbolic model checking PARAM~\cite{hahn2010param} to generate their correspondent symbolic formulae. For features whose nodes have context constraints, we restrict PARAM not to resolve context variability, leaving context conditions parameterized in the formulae. In our approach, uncertainties related to context conditions are only resolved at runtime. Therefore, we obtain \emph{unique} formulae for reliability and cost properties in terms of the context parameters, instead of maximum and minimum property formulae.
For the features that do not require a context condition, 
PARAM results are the same for both maximum and minimum property analysis. Table~\ref{paramTable} summarizes the formulae for the different features of a CGM node. Note that, in the first column, $N_1$, $N_2$, and $N_x$ represent subtrees. In the second and third columns, $P_n$ encapsulates the reliability and usage profile (i.e., frequency of execution), $W_n$ represents the cost, and $C_n$ represents the context condition of subtree $n$. $C_n$ may assume values 1 (true) or 0 (false) to indicate whether the context holds or not at a given time. Note also that the symbolic formulae for $N_x$ (that represents the incompleteness in the system) yields a variable $OPT$ that is either 1 (true) or 0 (false) to render the existence of the node's resource (e.g., a sensor or component) at runtime.

Even though the formulae of all structures are in terms of context condition ($C_n$), this information is only mandatory when using the DM-annotation. In this sense, the formulae of a node with purely AND/OR-decomposition or an incomplete node ($N_x$), but no uncertainty involved, follows Table~\ref{paramTable}, only without the parameter $C_n$. Also, in spite of the formulae for OR-decomposition and DM-annotation being similar, the semantics behind each structure is very different, since DM represents non-determinism in the system. We should note that, for the purpose of space and clarity, Table~\ref{paramTable} shows the formulae for a binary DM operator but the formulae for more operands is obtained in a compositional way. For a single operand, one just needs to set $C_{n2}$ to zero.

From Table~\ref{paramTable}, one can note that the reliability and cost parameters of all CGM node's features grow following a geometric progression with respect to the number of subtrees. For AND/OR-decomposition and incomplete node $N_x$, the reliability and cost parameters grow with a common ratio 2 and 3, respectively, in the best-case scenario (i.e., when there are no context conditions involved). In the worst-case scenario (i.e., when all subtrees are context dependent), the common ratio is 3 for reliability and 4 for cost parameters. The parameters of DM-annotation grow following a geometric progression with common ratio 3 for reliability and 4 for cost.

To generate reliability and cost parametric formulae for the overall system, we exploit the symbolic formulae on Table~\ref{paramTable} and the tree structure of goal models in a compositional way while taking the uncertainty modeled in the CGM into account: probability formulae of smaller goal models are computed and then composed to obtain the formula of its corresponding larger goal model. We follow a recursive depth-first strategy to visit the tree structure of the CGM. Each node gets a respective symbolic formulae in terms of the formulae associated with its sub-nodes. Leaf nodes get atomic formulae that are returned to rewrite their parents' formulae. Therefore, by the time the rewriting terminates, the system overall formulae will be a composition of its subtrees formulae.

Algorithm~\ref{lst:DEC_NODE} presents the compositional approach to build a parametric formulae. It starts from $node$, which is a local or the root goal of the CGM, and for which the parametric formula should be built. Line\,1 stores the subtree nodes of $node$ in the list $decNodes$. Line\,2 fetches the decomposition type $decType$, either AND or OR, of $node$. Line\,3 uses $dmAnnot$ to store the fetched DM-annotation of $node$. Line\,4 stores the context information of $node$ in $ctxAnnot$. Line\,5 calls function $getForm$, which returns in $nodeForm$ the parametric symbolic formula of $node$ considering the decomposition type ($decType$) and DM-annotation ($dmAnnot$) according to the symbolic formulae listed in Table~\ref{paramTable}.

\begin{algorithm}
{\footnotesize
 \KwIn{A node, either a root or a local goal}
 \KwResult{Parametric Symbolic formula of the node}
 \caption{composeNodeForm(Node node)}
 \label{lst:DEC_NODE}

 List [] decNodes $\leftarrow$ getDecomposition(node)\;
 DecType decType $\leftarrow$ getDecType(node)\;
 String dmAnnot $\leftarrow$ getDecisionMakingRule(node)\;
 String ctxAnnot $\leftarrow$ getContextInfo(node)\;
 String nodeForm $\leftarrow$ getForm(decType, dmAnnot, node)\;
 \ForEach{subNode in decNodes}{		
 	String subNodeId $\leftarrow$ getId(subNode)\;
    String subNodeForm $\leftarrow$ composeNodeForm(subNode)\;
	replaceSubForm(nodeForm,subNodeForm,subNodeId)\;
 }
 \If{isLeafTask(node)}{
   nodeForm $\leftarrow$ getParamForm()\;
 }
 \Return{nodeForm}\;
 }
\end{algorithm}

Accordingly and following a depth-first strategy, each $subNode$ of the goal tree is traversed through a recursive call to $composeNodeForm$, which produces a $subNodeForm$ that replaces its corresponding $ID$ symbol in $nodeForm$ (lines 6 to 9). Whenever the recursive approach finds a leaf task, the algorithm builds a parametric MDP of the task according to Listing~\ref{lst:PRISM_EXAMPLE}, and uses symbolic model checking to retrieve the task reliability and cost formulae. Finally, line 14 returns the parametric symbolic formula for the goal $node$.

\begin{figure}
\centerline{\includegraphics[width=1.0\columnwidth]{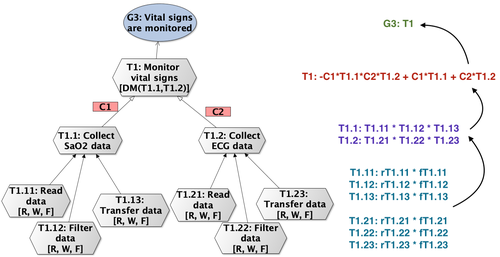}}
\caption{Example of Composing a Reliability Formula.}
\label{fig:paramExample}
\vspace{-1em}
\end{figure}

Fig.~\ref{fig:paramExample} exemplifies how the algorithm works for a fraction of the BSN's goal model to generate and compose a reliability formula. Since G3 has no specific feature, its reliability will be the same as its subtree ``T1: monitor vital signs". T1 has a DM-annotation, therefore its reliability follows the symbolic fomulae defined in Table~\ref{paramTable} for DM-annotations. Subtrees $T_{1.1}$ and $T_{1.2}$ have both an AND-decomposition, thus the reliabilities of each subtrees are multiplied to obtain the reliabilities of each, $T_{1.1}$ and $T_{1.2}$ (note that $T_{1.1}$ and $T_{1.2}$ have no further context conditions so that the corresponding variables $C_n$ are set to 1). Finally, the leaf nodes have their reliability retrieved by PARAM, in which $rT_i$ and $fT_i$ represent the reliability and execution frequency of leaf node~$i$. Similarly, cost formulae are generated by the algorithm.

The variability of the CGM contexts at runtime will create a specific induced Discrete Time Markov Chain (DTMC) to resolve any non-determinism at runtime. Accordingly, the system overall formulae vary as well. Table~\ref{tab:adversaries} shows the different context conditions the system may face during execution, and the reliability formula for each of them. We should note that the formula is specified in terms of \emph{reliability}, \emph{frequency of execution}, and \emph{context condition}, which are all uncertainties parameterized in the CGM. In the case of a cost formula, it would also be defined in terms of \emph{cost} of nodes. The computation of the reliability and cost formulae is used to evaluate the trustworthiness of the 
system based on the dynamic variance of (i)~reliability, (ii)~execution frequency, (iii)~cost, and (iv)~context conditions as reflected in the CGM. 
Also, by mapping the formulae parameters, one is able to guide the synthesis of adaptation policies by retrieving valid combinations of nodes to be executed at a given time. Next, we detail the runtime phase of our approach.

\begin{table}[htb]
\renewcommand{\arraystretch}{1.4}
\caption{Context Variability and Associated Formula}
\label{tab:adversaries}
\centering
\begin{tabular}{@{}lll@{}}
\hline
\textbf{C1} & \textbf{C2} & \textbf{Reliability formula for G3}\\
\hline
1 & 1 & \multirow{4}{*}{}$-rT_{1.11}fT_{1.11}rT_{1.12}fT_{1.12}rT_{1.13}fT_{1.13}*C_1$\\ & & $* rT_{1.21}fT_{1.21}rT_{1.22}fT_{1.22}rT_{1.23}fT_{1.23}*C_2$\\ & & $+ rT_{1.11}fT_{1.11}rT_{1.12}fT_{1.12}rT_{1.13}fT_{1.13}*C_1$\\ & & $+ rT_{1.21}fT_{1.21}rT_{1.22}fT_{1.22}rT_{1.23}fT_{1.23}*C_2$ \\
1 & 0 & $rT_{1.11}fT_{1.11}*rT_{1.12}fT_{1.12}*rT_{1.13}fT_{1.13}*C_1$\\
0 & 1 & $rT_{1.21}fT_{1.21}*rT_{1.22}fT_{1.22}*rT_{1.23}fT_{1.23}*C_2$\\
\hline
\end{tabular}
\end{table}

\subsection{Policy synthesis and enactment}

\subsubsection{Adaptation policy synthesis} \label{subsec:PolicySynth}

In our approach, the adaptation policy synthesis relies on assembling a set of adaptation goals and a set of possible actions. It is guided by parametric formulae generated through the aforementioned translation method, thereby assuring goal-oriented trustworthy behavior at uncertain execution scenarios. Systematically, the synthesis is composed by three steps: policy goals definition, contexts and actions identification, and policies enactment. 

\paragraph{Policy goals definition}
Based on the augmented CGM, the requirements engineer should define a set of properties (e.g., cost, reliability, performance) that the managed system must satisfy during runtime and that attends to the purpose of the policy. In our BSN example, we propose an energy-saving policy that must deal with the reliability-battery consumption trade-off. Therefore, we list two properties: 
(1)~the probability of successfully achieving G1 must be within 90\% with a 2\% error margin;  
(2)~the battery consumption (cost) for achieving G1 must be within 0.47W with a 2\% error margin. 
Thus, the policy goal definition consists on combining the elicited properties into logic propositional sentences that semantically attends to the purpose of the policy. In our example, '(1) AND (2)' shall be continuously satisfied. The policy goals are used as set points for guiding the desired behavior of the managed system.

\paragraph{Actions and contexts identification}

Once the properties and policy goals to be achieved are defined, the domain expert must identify a set of contexts that shall be monitored by the managing system, and a set of possible actions that can be performed to modify the managed system's behavior. This step is highly dependent on the overall architecture, the managed system's knobs and sensors, and the context's sensors. For example, BSN operates with a publish/subscribe architecture and its managing system monitors and act upon the managed system through message exchange. %

The context in which the system operates may be prone to high variability, so it requires flexible adaptation goals sensitive to contextual information. Therefore, we should note that the domain expert shall employ context operationalization techniques to enhance contextual knowledge when creating the policies. In our BSN example (see Fig.~\ref{fig:ND-CGM}), the contexts $C_1$ to $C_6$ regard the availability and data correctness of SaO2, ECG, temperature, and blood pressure sensors. At runtime, the components responsible for the leaf tasks (collect, filter, and send) can constantly inform the managing system whether the contexts are active or not (e.g., whether a sensor is available).

Actions on the other hand are susceptible to the system adaptation capabilities. Each action has a side effect which is accounted in the properties evaluation before the definition of eligible policies for achieving the adaptation goals. For example, each node in the BSN has an execution frequency that is directly proportional to the task's reliability and cost, in which the managing system can act upon to achieve the adaptation goals at runtime.

\subsubsection{Policies enactment}\label{subsec:polEnact}

Once the policies have been synthesized, they are taken to runtime as a set of possible actions that will be exercised via the parametric formulae through an exhaustive search for the current runtime situation. These actions guide the behavior of the managed system to achieve the adaptation goals in face of uncertainty.

\begin{figure}[t]
\centerline{\includegraphics[width=1\columnwidth]{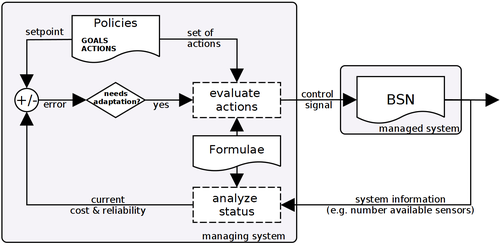}}
\caption{Control Feedback Loop for the BSN Example.}
\label{fig:control_loop}
\end{figure}

Specifically, as illustrated in Fig.~\ref{fig:control_loop}, the managing system enforces at runtime the satisfaction of properties by continuously monitoring the tasks' internal properties (cost, reliability, and frequency) and the active contexts (in terms of availability of sensors), which all undergo an analysis process using the parametric formulae. Then, the difference between the actual cost/reliability status and the set point defined by the policy is computed, and may trigger distinct actions for execution. Since the uncertainty parameters are expressed in the formulae originating from the augmented CGM, they are resolved whenever a set of actions is executed to cope with transgressed reference values. For instance for the BSN (Fig.~\ref{fig:ND-CGM}), we exercise the root goal G1 with the tamed and untamed availability of sensors using the energy-saving policy described throughout this section, and obtain the curves plotted in Fig.~\ref{fig:res_asen}.
 
\section{Evaluation}
\label{sec:experiments}

We evaluate our approach on the BSN prototype\footnote{The BSN implementation and data used in this section are available at GitHub: \url{https://github.com/rdinizcal/bsn_Seams2019}.} by means of a Goal-Question-Metric (GQM) methodology~\cite{Basili:1994}. 
The evaluation compares two versions of the BSN, one with a controller with untamed uncertainty (without our approach), and the other one with a controller with tamed uncertainty (with our approach). We do not compare with a version of the BSN that has no controller since it is not a self-adaptive system. %
Therefore, we aim at analyzing if the application of our approach in the BSN is able to assist the adaptation strategies with meeting its reliability and cost goals under the presence of uncertainties.  
Table~\ref{tab:gqm} shows the goal, question, and metrics for the evaluation.

\begin{table}[hbtp]
\caption{Goals, Questions, and Metrics of the Evaluation.}
\label{tab:gqm}
\centering
\footnotesize
\vspace{-1.25em}
\begin{tabular}{|p{0.35\textwidth}|p{.09\textwidth}|}
\multicolumn{1}{l}{}\\\hline
\multicolumn{2}{|c|}{\textbf{Evaluation Goal: Uncertainty impact on SAS assurance process}}\\\hline
\multicolumn{1}{|c|}{\texttt{Question}}& \multicolumn{1}{c|}{\texttt{Metrics}}\\\hline

\texttt{How does our approach enhance reliability and cost when taming uncertainty coming from:  system itself (1a), system goals (1b), and environment (1c)?} & \texttt{$e_{r}$ and $e_{c}$} \\\hline

\end{tabular}
\end{table}

In Table~\ref{tab:gqm}, $e_r$ and $e_c$ are the enhancements of reliability and cost, respectively. By enhancements we quantitatively measure the gain in achieving the adaptation goals (i.e., reliability and cost) that our approach promotes to the adaptation in the presence of uncertainties compared to not using our approach. These metrics are evaluated with reliability and cost data collected over time, in BSN, at scenarios with different classes of uncertainty using our generated parametric formulae. We consider the impact of the adaptation policies devised from the parametric formulae where the classes of uncertainty are either \emph{untamed} or \emph{tamed}, and derive the relation between the average distance $d$ to the set point for the \emph{untamed} ($d_{untamed}$) and \emph{tamed} ($d_{tamed}$) curves, where $ d = \dfrac{\sum|x_i - x|}{n}$ and $e_{x} = \dfrac{d_{untamed}}{d_{tamed}}$, where $x=r$ (for reliability) or $x=c$ (for cost).%

\subsection{Experimental Evaluation in the BSN}

Our experimental evaluation consists of exploring distinct classes of uncertainty scenarios, their impact on BSN, and how our approach enhances reliability and cost during runtime. Thus, the evaluation aims at confirming how our approach leads engineers to build controllers that tame distinct classes of uncertainty from an end-to-end goal-oriented perspective. Therefore, we perform two executions for each scenario that raise distinctions in the BSN performance: (1) controlled behavior with untamed uncertainty and (2) controlled behavior with tamed uncertainty. %
We compare how well the devised adaptation policies, guided by the generated formulae and the goal model either with tamed or untamed uncertainty, assures that the managed system performs with respect to its goals.

\subsubsection{Experimental Setup}\footnote{Configuration for experiments: CPU 4x Intel(R) Core(TM) i7-5500U CPU @2.40GHz, 8075MB RAM, Ubuntu 16.04.3 LTS, GNU C Compiler version 5.4.020160609, hard drive ATA Corsair Force LE.}
The experiments were ran in the BSN, implemented with the OpenDaVINCI~\cite{opendavinci} framework. To evaluate the formulae at runtime we use the Lepton expression parser library~\cite{lepton}.

For all scenarios, we employ an energy-saving policy, such as  Sec.~\ref{subsec:PolicySynth}, that aims at balancing reliability and cost by exploring the relation between the components' execution frequencies 
:``the root goal G1 must be achieved with ($90$ $\pm$ $2$)\% reliability and ($0.47$ $\pm$ $0.01$)W cost; the policy actions are: \emph{adjust sensor's sampling rate} and \emph{adjust central hub's execution frequency}.

\subsubsection{Experimental scenarios and results}

\textbf{\\Class of uncertainty: System itself.} In this scenario, we explore the direct impact of messages waiting to be read by the central hub on the system's goal for reliability and cost, regarding the evaluation goal~\textbf{(1a)}. Here, we simulate the degrading of the tasks' reliability related to the delays on processing messages due to unexpected number of patients using the system. The patients are simulated by modules that flood the central hub communication channel. 

\begin{figure*}[htp]    
    \centering
    \begin{subfigure}[t]{.20\textwidth}\centering
        \includegraphics[width=1\columnwidth]{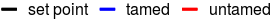}
    \end{subfigure}
    \\
     \begin{subfigure}[t]{.30\textwidth}\centering
        \includegraphics[width=1\columnwidth]{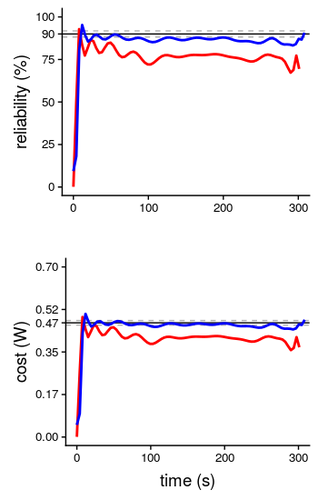}
        \caption{Uncertain Number of Users.}
        \label{fig:res_ch}
    \end{subfigure}
    \quad
    \begin{subfigure}[t]{.30\textwidth}\centering
        \includegraphics[width=1\columnwidth]{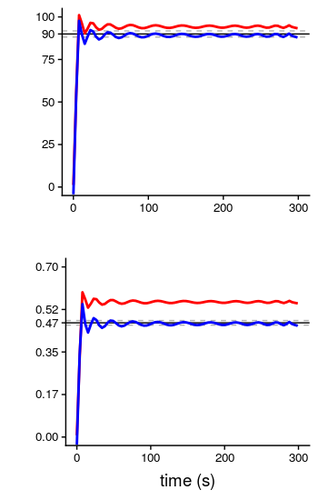}
        \caption{Uncertain Sampling Rate of Sensors.}
        \label{fig:res_frq}
    \end{subfigure}
    \quad
    \begin{subfigure}[t]{.30\textwidth}
        \includegraphics[width=1\columnwidth]{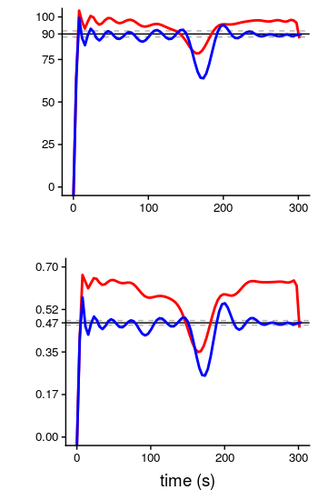}
        \caption{Uncertain Availability of Sensors.}
        \label{fig:res_asen}
    \end{subfigure}
    \caption{Reliability and Cost Behavior for Different Classes of Uncertainty.}
    \vspace{-1em}
\end{figure*}

The untamed behavior depicted in Fig.~\ref{fig:res_ch}, is due to the managing system's unawareness of the reliability degrading effect in the central hub, hence, we simulate the system engineer estimating and setting static reliability values in the managing system. This results in the reliability drift observed during the execution, as long as new modules simulating patients are instantiated and the system cannot repair due to the difference between estimated central hub's reliability and the actual value. 
The tamed behavior also shows a drift regarding the instantiations of new modules simulating patients, however, it maintains the system within the set point boundaries, confirming that the G1 is satisfied (see Fig.~\ref{fig:res_ch}). %
The enhancement values for this evaluation scenario are $e_{r} = 2.66$ and $e_{c} = 3.36$. Thus, our approach enhanced the reliability by 2.66 and the cost by 3.36 relative to the set points compared to the untamed setting for this scenario.

\textbf{Class of uncertainty: System goals.} To investigate the evaluation of this class of uncertainty~\textbf{(1b)}, we consider a scenario in which the understanding of the system goals is inaccurate, such as how the reliability and cost properties are related to the patient's profile, which could demand distinct system goals according to the patient's needs. For this purpose, we simulate a scenario in which the doctor configures the BSN sensors' sampling rates with values based on the patient's profile. For example, consider a cardiac patient who needs careful attention to heart rate. To optimize the battery consumption, one could increase the sampling rate of the ECG and decrease the rate of the thermometer. However, a mistaken set up, due for example to uncertain needs and profile of a patient, could lead to unnecessary battery consumption and misguided adaptations with respect to the reliability and cost properties.

The untamed behavior depicted in Fig.~\ref{fig:res_frq} is due to uncertainty related to the sampling rate associated with the patient profile. Thus, we simulate a situation where a system engineer sets the managing system with a static sampling rate for each sensor that does not correspond to the needs and profile of the patient, for whom the BSN is operating. The managing system then prompts the adaptation to the incorrect set point resulting in the shifted behavior. On the other hand, the tamed behavior adapts to the required set point boundary in the first moments of the execution, assuring that the goal G1 is satisfied (see Fig.~\ref{fig:res_frq}).
The corresponding enhancement values for this evaluation scenario are $e_{r} = 3.04$ and $e_{c} = 9.30$.

\textbf{Class of uncertainty: Environment.} In this scenario, we explore how the variation of available sensors impacts the system's goal of reliability and cost, thus investigated evaluation goal~\textbf{(1c)}. We simulate a local autonomous recharging mechanism for each sensor that deactivates the sensor when the battery level is less than 2\%, and reactivates it when the level achieves at least 90\%. This results in an unpredictable variation of the number of available sensors at runtime.

The untamed behavior depicted in Fig.~\ref{fig:res_asen} is due to the managing system's uncertainty about the number of available sensors. Thus, we simulate a situation where the managed system is unable to communicate the availability of its sensors. In face of this uncertainty, the managing system's engineer estimates that all sensors are available. %
The disparity between the estimated and the actual number of available sensors explains the shifted cost and reliability curves in the untamed execution. 
In contrast to the untamed behavior, the tamed behavior evidences that the adaptations performed by the managing system assures that the goal G1 is satisfied within the adaptation goals requirements (see Figure~\ref{fig:res_asen}). 
The corresponding enhancement values for this evaluation scenario are $e_{r} = 1.98$ and $e_{c} = 4.09$.

We should note the situation around 180s, when three sensors become unavailable at the same instant and the managing system could not find a strategy to satisfy the adaptation goals. A few seconds later, one sensor was recharged and became available so that the managing system could cope with the situation and regain system stability. While the tamed controller is able to reach steadiness, the untamed controller still faces a drop in reliability and in cost close to the 300s.

\subsubsection{Summary}

The experimental results show that our approach provides an effective way to tackle three classes of uncertainty: (1a)~system itself, (1b)~system goals, and (1c)~environment, with enhancement gains ($e_{x}$) ranging from 1.98 to 3.04 for reliability and 3.36 to 9.30 for cost. This means that, in the worst case, our approach prompted an enhancement of the BSN to achieve its adaptation goals in face of uncertainties of almost twice as much; and in the best-case scenario of more than nine times as much. Moreover, the runtime verification seems to be quite affordable, considering a system like BSN whose cost formula is generated in the mean time of 0.085s with 60 parameters at design time, 22 KB of storage, and evaluated within 0.020s at runtime.

\subsection{Threats to validity}

\noindent\textbf{Construct validity.} The major threats here are the correctness of the implementation of the BSN and of the proposed approach. The BSN has been thoroughly tested as part of previous work~\cite{Rodrigues:learning}. Concerning our approach, at least two authors of this paper reviewed the implementation and checked the plausibility of the evaluation results based on the experience they have with the BSN.

\noindent\textbf{Internal validity.} Our approach showed itself effective and efficient in the evaluation. Although we comprehensively deal with uncertainties classes, unveiling \textit{all} sources of uncertainty involved in a system's operation is inherently non-deterministic, which represents a threat to any assurance process. Moreover, our policy strategies were created manually, which might represent a scalability threat in complex scenarios with multiple uncertainties combined as well as in catastrophic scenarios where highly rare events have to be taken into account. 

\noindent\textbf{External validity.} Although our approach is platform independent, we do reckon the limitation of the evaluation since it was applied in the specific case of the BSN. Further evaluation must be performed to generalize the results. Despite all efforts to implement the BSN with new features to tame uncertainty, further study must be done to verify the applicability in real-world scenarios with multiple uncertainties combined.

\section{Related Work}
\label{sec:relatedwork}

Regarding the modeling and verification approaches for SAS under uncertainty, Filieri \textit{et al.} \cite{filieri2016supporting} present a framework to overcome model checking scalability issues by generating runtime symbolic expressions of system requirements from rewarded Discrete-Time Markov Chains, while accounting for uncertainty in the managed system. Weyns and Iftikhar\cite{weyns2016model} propose a modular approach for decision making that supports changing goals at runtime. They combine distinct models for each relevant quality of the system with runtime simulation of the models to select an adaptation option that satisfies the system goals. Bencomo and Belaggoum\cite{bencomo2013supporting} map goal models onto Dynamic Decision Networks (DDNs) to provide a principled approach to make rational decisions in the face of uncertainty within changing environments. Cailliau and Lamsweerde\cite{cailliau2017runtime} use goal models at runtime to support adaptation aiming at increasing the actual satisfaction rate of probabilistic system goals in spite of environment changes. They use probabilistic $LTL_3$ monitors to continuously evaluate goals satisfaction and find most appropriate countermeasures at runtime. In our work, we augment a goal-oriented approach into modeling SAS within multiple sources from different classes of uncertainty, from which we automatically generate verifiable runtime models parameterized with these uncertainties. Similar to Filieri \textit{et al.}\cite{filieri2016supporting}, we use symbolic model checking to overcome model checking scalability issues at runtime. But, our symbolic formulae are also able to guide the synthesis of adaptation policies for self-adaptation.

Regarding the support of uncertainty in the assurance process of an end-to-end methodology, Ghezzi \textit{et al.}\cite{ghezzi2013managing} present ADAM, a model-driven framework conceived to support the development and runtime operation of SAS, aiming at mitigating uncertainty concerning response time and faulty behavior. Calinescu \textit{et al.} \cite{QoSMOS} propose QoSMOS, a framework to develop service-based systems that achieve their Quality of Service (QoS) requirements through dynamically adapting to changes in the system state, environment, and workload. In \cite{calinescu2017entrust}, Calinescu \textit{et al.} propose the ENTRUST methodology to provide assurance evidence, cases, and arguments for SAS at design- and runtime while supporting internal and environmental changes. In our work, we go further on addressing different classes of uncertainty, since we tame three classes with focus on five sources of uncertainty: (i)~future parameter value, and (ii)~incompleteness, regarding uncertainty in the system itself; (iii)~specification of goals, regarding uncertainty in system goals; (iv)~execution context, and (v)~noise in sensing, regarding environmental uncertainty.

Among the works that focus on synthesizing adaptation policies, Su \textit{et al.}\cite{su2016iterative} use a parametric MDP to apply the value iteration method and select a confidently optimal adaptation policy at runtime with focus on a trade-off among accuracy, data usage, and computational overhead. C\'amara \textit{et al.}\cite{camara2018mosaico} and Moreno \textit{et al.} \cite{moreno2015proactive} use the PRISM language to model SAS as MDPs and synthesize adaptation policies. They both use the probabilistic model checker PRISM, but while C\'amara \textit{et al.}\cite{camara2018mosaico} focus on a design-time approach to synthesize optimal repertories considering uncertainty, Moreno \textit{et al.} \cite{moreno2015proactive} use PRISM at runtime to synthesize policies that maximize an expected accumulated utility over a finite horizon. In our work, we synthesize adaptation policies by means of our design-time generated symbolic formulae parameterized with \emph{multiple classes of uncertainty}. In this sense, we avoid the model exploration that is required by traditional model checking techniques while supporting various types of uncertainty.

\section{Conclusion and Future Work}
\label{sec:conclusion}
In this work, we showed an assurance process for trustworthy SAS that is able to comprehensively and systematically tame various sources from different classes of uncertainty based on Goal-Oriented Requirements Engineering (GORE) and that covers the design- and runtime of the SAS. At design time, we augment goal modeling with uncertainties. From the resulting goal models, we compositionally generate reliability and cost parametric formulae provided with parameterized uncertainties. These formulae are then used to synthesize adaptation policies and, at runtime, to automatically evaluate the fulfillment of SAS goals taking different classes of uncertainty into account while supporting self-adaptation. 
The evaluation on the BSN shows that our approach is effective in taming different classes of uncertainty and efficient in performing symbolic model checking of SAS parameterized with multiple classes of uncertainty, assuming a reasonable number of parameters.

As future work, we aim for an approach that supports more scalable and automated synthesis of adaptation policies, while quantitatively modeling uncertainty (e.g., with probability distributions). In addition, we plan to provide means to cope with the combinatorial possibilities in the DM-annotation in the goal model as well as how rare events such as catastrophic scenarios could be taken into account. Finally, we plan to evaluate our approach with multiple SAS exemplars.

\section*{Acknowledgment}

The authors express their utmost gratitude to L\'{e}o Morais and Gabriel Levi (UnB/Brazil) for implementing the extensions of the BSN, as well as to Jo\~{a}o Henrique Pimentel (UFRPE/Brazil) and Leandro Bergmann (UnB/Brazil) for their support in the implementation of GODA's new version in PiSTAR. This study was financed in part by the CAPES-Brasil -- Finance Code 001, through LEAPaD/PROCAD research project funding for Gena\'{i}na and CAPES scholarships for Ricardo and Gabriela. Gena\'{i}na also thanks CNPq for partial support under grant number 306017/2018-0.

\balance

\bibliographystyle{IEEEtran}
\bibliography{seams2019}

\end{document}